\documentclass[12pt]{article}
\usepackage{amsmath}
\usepackage{amssymb}
\usepackage{graphicx}
\usepackage{psfrag}
\usepackage{epsf}
\usepackage{enumerate}
\usepackage{sectsty}

\usepackage{lscape}
\usepackage{booktabs}
\usepackage{setspace}
\usepackage[margin=1in]{geometry}
\newcommand{\E}{\mathbb{E}}

\newcommand{\ci}{\mathrel{\text{\scalebox{1.07}{$\perp\mkern-10mu\perp$}}}}

\newcommand{\nmb}{\text{NMB}}
\newcommand{\cea}{\text{CEA}}
\newcommand{\bL}{\textbf{\textit{L}}}
\newcommand{\bV}{\textbf{\textit{V}}}
\newcommand{\bW}{\textbf{\textit{W}}}
\newcommand{\Pro}{\text{P}}
\sectionfont{\fontsize{15}{15}\selectfont}

\newcommand{\blind}{1}

\begin{document}

\def\spacingset#1{\renewcommand{\baselinestretch}%
{#1}\small\normalsize} \spacingset{1}


\if1\blind
{
  \title{\bf \Large Net benefit separation and the determination curve: a probabilistic framework for cost-effectiveness estimation}
  \author{\normalsize Andrew J. Spieker$^{1}$,\\ \normalsize Nicholas Illenberger$^{2}$,\\ \normalsize Jason A. Roy$^{3}$,\\ \normalsize Nandita Mitra$^{2}$\\}
  \maketitle
} \fi

{\small
\noindent%
{$^1$} Vanderbilt University Medical Center; Department of Biostatistics (\textit{andrew.spieker@vumc.org}).\\
{$^2$} University of Pennsylvania; Department of Biostatsitics, Epidemiology, and Informatics.\\
{$^3$} Rutgers University; Department of Biostatistics and Epidemiology.
}

\begin{abstract}
Considerations regarding clinical effectiveness and cost are essential in comparing the overall value of two treatments. There has been growing interest in methodology to integrate cost and effectiveness measures in order to inform policy and promote adequate resource allocation. The net monetary benefit aggregates information on differences in mean cost and clinical outcomes; the cost-effectiveness acceptability curve was then developed to characterize the extent to which the strength of evidence regarding net monetary benefit changes with fluctuations in the willingness-to-pay threshold. Methods to derive insights from characteristics of the cost/clinical outcomes besides mean differences remain undeveloped but may also be informative. We propose a novel probabilistic measure of cost-effectiveness based on the stochastic ordering of the individual net benefit distribution under each treatment. Our approach is able to accommodate features frequently encountered in observational data including confounding and censoring, and complements the net monetary benefit in the insights it provides. We conduct a range of simulations to evaluate finite-sample performance and illustrate our proposed approach using simulated data based on a study of endometrial cancer patients.
\end{abstract}

\vfill

\noindent%
{\it Keywords:} Censoring, Confounding, Cost-effectiveness, Observational, Policy

\newpage
\spacingset{1.05} 

\section{Introduction}
Policy decisions are often informed by aggregate information on cost and clinical effectiveness. Several measures have been previously proposed to compare the cost-effectiveness of two interventions. Arguably, the simplest of these measures is the incremental cost-effectiveness ratio (ICER), given by the difference in mean cost divided by the difference in mean clinical measure \cite{Polsky1997, Willan2006, willan1996confidence}. When the ICER lies below a pre-determined willingness-to-pay (WTP) threshold, the experimental treatment is deemed cost-effective. The net monetary benefit (NMB) is an equivalent measure to the ICER that was proposed to avoid a singularity in the denominator, as well as known challenges associated with interval estimation of ratio parameters \cite{Briggs1997, Heitjan19992}. The NMB has been used to compare cost-effectiveness in many recent studies \cite{Leppert2018, Liao2017, Zhang2017, Maru2015, Shafrin2018}.

The cost-effectiveness acceptability (CEA) curve was proposed as a probabilistic representation of strength of evidence regarding NMB; its function is to elucidate how conclusions vary with changes in the WTP \cite{Lothgren2000}. The CEA is defined as the proportion of positive bootstrapped estimates of the NMB; it is sometimes loosely referred to as the "probability of cost-effectiveness," \cite{Briggs2005, Fenwick2004, Fenwick06} but more closely resembles a frequentist \textit{p}-value \cite{Hoch2008}. The CEA has been used as a primary tool to advocate policy changes. Delaney et al., for instance, utilize a CEA as evidence in support of initial endoscopy in dyspepsia patients \cite{delaney2000cost}, and mental health studies have commonly included CEAs for decision-making purposes in various interventions, sometimes under the name "probabilistic sensitivity analysis" \cite{Ahern2018, Hunter2014}. However, Barton et al. note that the CEA, despite its widespread use, does not alone provide a path for optimal decision-making as it does not provide insights into the extent of cost-effectiveness \cite{Barton2008}.

Some have proposed alternative graphical tools to the CEA in response to the challenges associated with interpreting it as a probability of cost-effectiveness within the frequentist paradigm. One approach, proposed by Heitjan et al., was to re-frame the NMB and acceptability curve within a Bayesian paradigm so that a probabilistic interpretation would be warranted \cite{Heitjan1999, Heitjan2004}. Hoch and Blume present a likelihood ratio approach based on a pre-specified alternative hypothesis \cite{Hoch2008}. These approaches, like the CEA, are still guided by inference and strength of evidence regarding the NMB, quantifying either the degree of belief regarding the mean cost-effectiveness \textit{a posteriori} (Bayesian methods), or the strength of evidence in support of a positive NMB (likelihood methods).

While the CEA provides insights into sensitivity of conclusions regarding mean clinical/cost outcome combinations to fluctuations in the WTP, a probabilistic measure of cost-effectiveness \textit{magnitude}---i.e., a parameter that quantifies the frequency with which treatment produces more desirable clinical/cost outcome combinations---can summarize the distribution of the cost/clinical outcomes as a whole in a way that complements insights from mean differences. In this paper, we define the \textit{net benefit separation} (NBS) as the probability that an observation from a hypothetical population in which all patients receive treatment has greater individual net benefit than an observation from a hypothetical population in which all patients receive the control. This gives rise to a visualization tool in which the NBS is plotted over a range of willingness-to-pay thresholds, which we term the \textit{cost-effectiveness determination} (CED) curve. In this paper, we will further demonstrate how weighting and standardization can be used to estimate the NBS and its corresponding CED curve when outcomes are right-censored and the data are subject to confounding.

The remainder of this paper is organized as follows. In Section 2, we provide a review of the net monetary benefit and its associated cost-effectiveness acceptability curve. In Section 3, we define the net benefit separation and its associated determination curve. In Section 4, we propose an estimation procedure to estimate net benefit separation using observational data. In Section 5, we present results from a simulation study in order to understand the finite-sample properties of our methods. In Section 6, we present an illustration of these methods using simulation-based data from a prior study of endometrial cancer patients \cite{spieker2017causal}. We conclude with a discussion of our findings in Section 7.

\section{Review of the Net Monetary Benefit and Acceptability Curve}

\subsection{The net monetary benefit}

Let $i = 1, \dots, N$ index independently sampled study participants, and let $A_i$ denote binary treatment. Let $S_i$ denote the clinical measure (i.e., the outcome that determines the treatment's effectiveness), and $Y_i$ the cost. For example, $S$ could be systolic blood pressure reduction or survival time, and $Y$ could be initial treatment cost or total medical cost billed through some time range. The net monetary benefit (NMB) is defined as follows:
\begin{eqnarray*}
	\nmb(\lambda) &=& \lambda(\E[S|A = 1] - \E[S|A = 0]) - (\E[Y|A = 1] - \E[Y|A = 0]).
\end{eqnarray*}
Here, $\lambda$ denotes the WTP and is defined as the maximum cost one would be willing to pay for a one-unit improvement in $S$ associated with treatment $A = 1$ (for example, \$50,000 for year of life saved). Note that $\nmb(\lambda) > 0$ indicates treatment $A = 1$ is more cost-effective on average, while $\nmb(\lambda) < 0$ indicates the reverse.

\subsection{The cost-effectiveness acceptability curve}

The CEA is defined according to the following procedure:
\begin{enumerate}
\item Define some range $\Lambda$ of possible willingness-to-pay thresholds.
\item For $k = 1, \dots, K$:
\vspace{-2mm}
\begin{itemize}
\item Draw a full-sized bootstrap replicate from the set $\lbrace A_i, S_i, Y_i \rbrace_{i = 1}^{N}$.
\item Compute $\widehat{\nmb}_k(\lambda)$ using the bootstrapped data for each $\lambda \in \Lambda$.
\end{itemize}
\vspace{-2mm}
\item For $\lambda \in \Lambda$, the acceptability curve is defined as:
$$\text{CEA}(\lambda) = K^{-1}\sum_{k = 1}^{K} 1(\widehat{\nmb}_k(\lambda) > 0).$$
\end{enumerate}
While sometimes stated as the probability of cost-effectiveness \cite{Briggs2005, Fenwick2004, jiaqili2018, Fenwick06, Lothgren2000}, this is more accurately described as one minus a one-sided bootstrapped \textit{p}-value of the lesser alternative that the control is more cost-effective on average. The limiting behavior of the CEA can be described as follows. First fix $N = n$: at $\lambda = 0$, clinical effects are not considered, and as $\lambda$ increases, the mean cost difference wanes in its influence on $\cea(\lambda)$. Now consider asymptotics in $N$, and suppose first that the difference in mean clinical outcome is nonzero (the non-degenerate case). As $N$ grows, the CEA converges point-wise to a step function:
\begin{eqnarray*}
\lim_{N, K \longrightarrow \infty} \cea(\lambda) &=& \begin{cases} 1 & \mbox{ for } \lambda > \frac{\E[Y|A = 1] - \E[Y|A = 0]}{\E[S|A = 1] - \E[S|A = 0]}  \\ 
0.5 & \mbox{ for } \lambda = \frac{\E[Y|A = 1] - \E[Y|A = 0]}{\E[S|A = 1] - \E[S|A = 0]} \\
0 & \mbox{ for } \lambda < \frac{\E[Y|A = 1] - \E[Y|A = 0]}{\E[S|A = 1] - \E[S|A = 0]} \end{cases}.
\end{eqnarray*}
If instead the difference in mean clinical outcome is zero, then
\begin{eqnarray*}
\lim_{N, K \longrightarrow \infty} \cea(\lambda) &=& \begin{cases} 1 \hspace{1mm} \forall \lambda & \mbox{ if } \E[Y|A = 1] < \E[Y|A = 0] \\
0.5 \hspace{1mm} \forall \lambda & \mbox{ if } \E[Y|A = 1] = \E[Y|A = 0] \\
0 \hspace{1mm} \forall \lambda & \mbox{ if } \E[Y|A = 1] > \E[Y|A = 0] \end{cases}.
\end{eqnarray*}
In the non-degenerate case, the limit of the CEA informs the value of $\lambda$ at which the treatments are equally cost-effective on average (i.e., at the ICER); equivalently, it informs the values of $\lambda$ at which the NMB is positive, regardless of magnitude. Figure 1 illustrates the CEA's limiting behavior when treatment produces higher mean cost and clinical benefit.

\begin{figure}[h!]
\centering
\includegraphics[width = 3.2in]{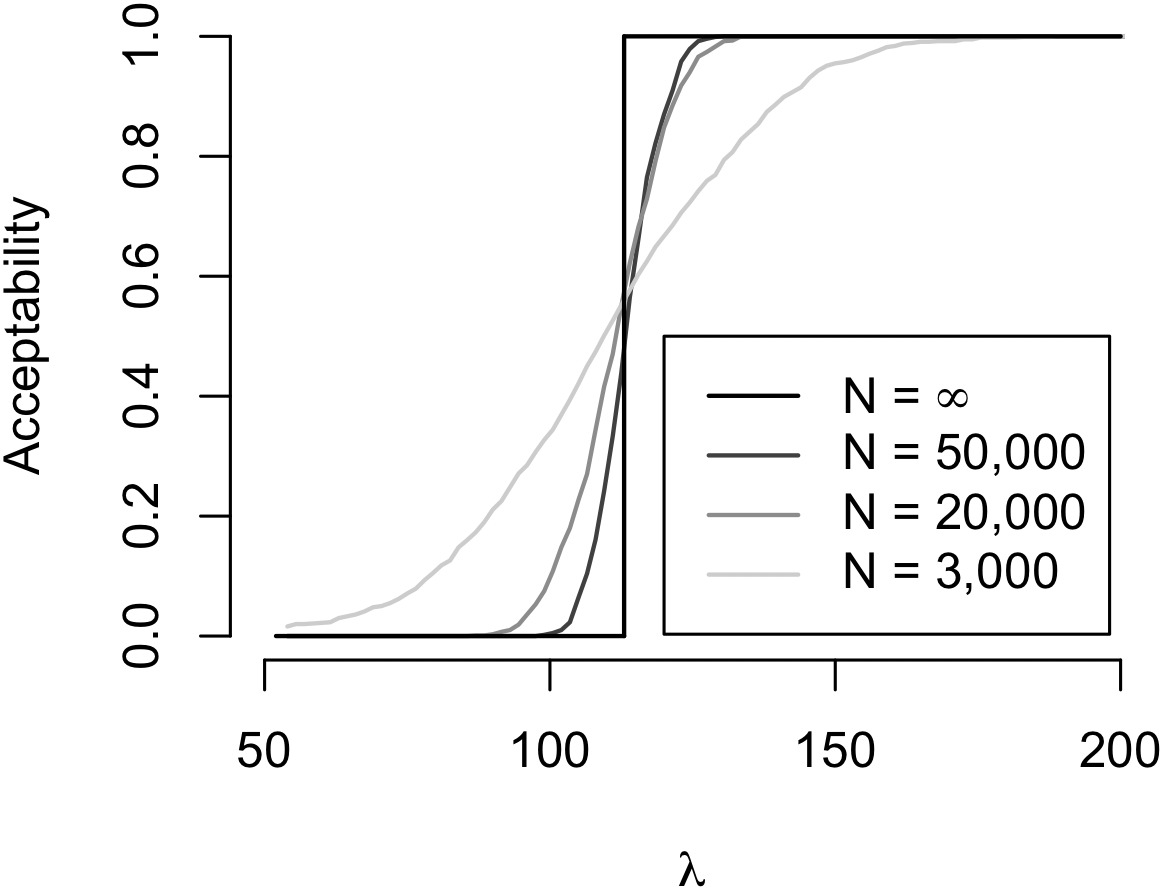}
\caption{Illustration of the limiting behavior of the cost-effectiveness acceptability curve. Here, the experimental treatment is more costly on average, but also yields greater mean clinical benefit (a prototypical setting in which a cost-effectiveness analysis would be warranted). Note that the step from zero to one in the limit of the CEA occurs precisely at the value of the ICER.}
\end{figure}

For a given $\lambda$, a higher CEA value indicates stronger support in favor of cost-effectiveness for treatment $A = 1$; however, values closer to zero do not distinguish between "absence of evidence for the experimental treatment's cost-effectiveness" and "evidence of cost-effectiveness for the control treatment." This challenge regarding lack of symmetry can be avoided by instead plotting the estimated NMB and confidence intervals across a range of $\lambda$. This graphical representation accomplishes the goal of the CEA (which is to provide insights into the sensitivity of conclusions regarding the value of the NMB to changes in $\lambda$) with the added advantage of depicting both the point and interval estimates.

\section{Net Benefit Separation: A Probabilistic Approach}

Recall that $(S, Y)$ denotes the clinical/cost outcome pair. Define $B(\lambda) = \lambda S - Y$ to be the individual net benefit (INB). Following the potential outcomes notation \cite{rubin1978}, let $(S^a, Y^a)$ denote the clinical and cost outcome pair under treatment $A = a$, and $B^a$ the INB under treatment $A = a$. One straightforward probabilistic metric of cost-effectiveness is defined at the subject specific level: $\theta^*(\lambda) = \Pro(B_i^1(\lambda) > B_i^0(\lambda))$. This can be described as the probability that the treatment is cost-effective for a random subject. Because only one potential outcome pair can be observed per subject, $\theta^*(\lambda)$ is not identifiable without imposing unrealistic and untestable assumptions (e.g., uncorrelatedness of $B_i^0(\lambda)$ and $B_i^1(\lambda)$).

We instead propose a measure of stochastic ordering of the marginal distributions under each hypothetical comparator treatment. To avoid notational confusion, let $B_a(\lambda)$ denote a randomly sampled value for $B(\lambda)$ under the setting in which everyone receives treatment $A = a$. We define the \textit{net benefit separation} (NBS) by $\theta(\lambda) = \Pro(B_1(\lambda) > B_0(\lambda))$. In essence, the NBS targets the area under the receiver operating characteristic curve, frequently used in the evaluation of diagnostic measures, and is so named as it takes on the value one only when the distribution of INBs under $A = 1$ stochastically dominates the distribution under $A = 0$, and zero if the reverse is true. The NBS takes on the value of one-half, for instance, when the treatment has null clinical and cost effects (Figure 2).\\

\begin{figure}[h!]
\centering
\includegraphics[width = 5.1in]{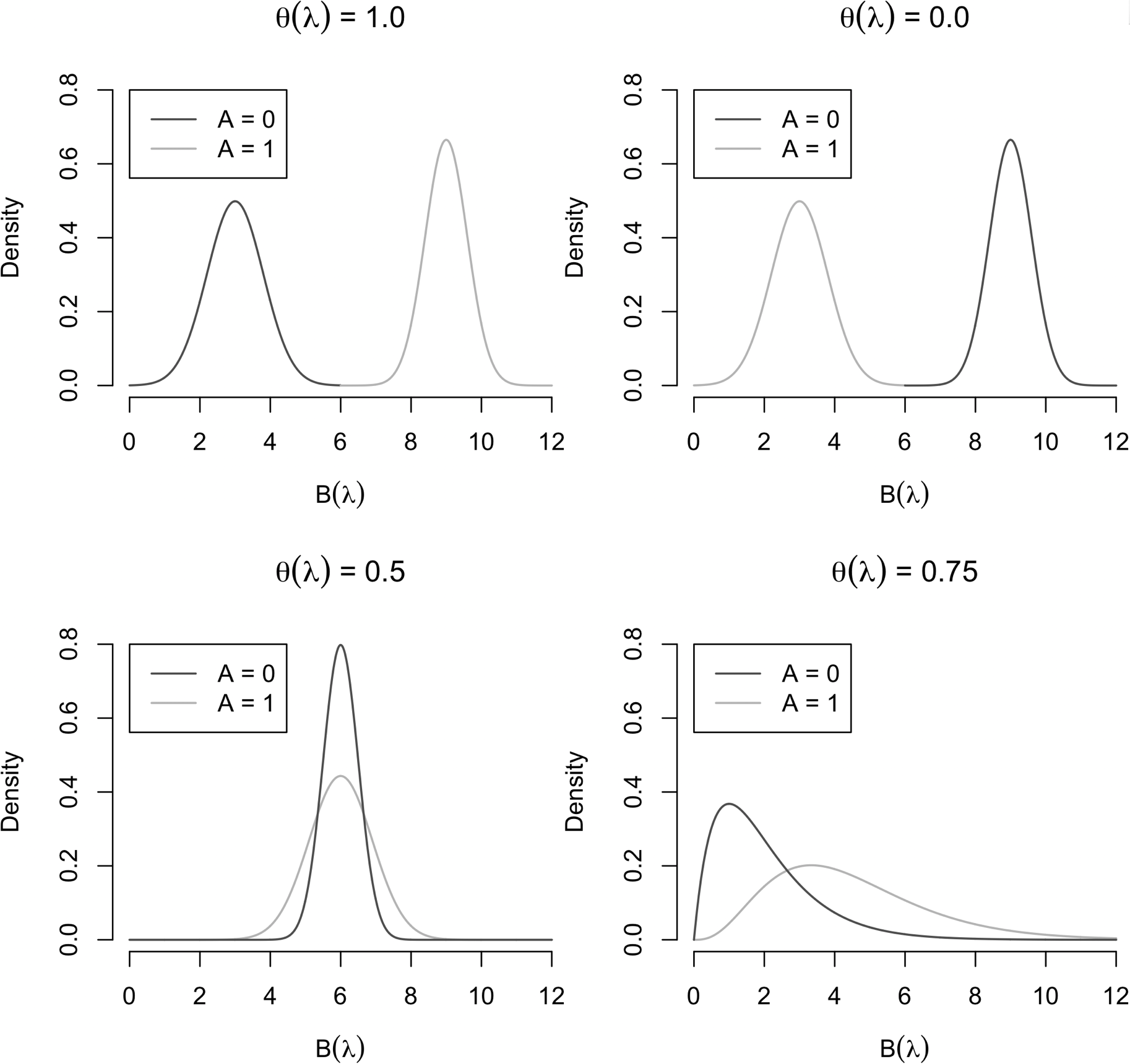}
\caption{Illustration of NBS under four scenarios. In the first (upper left), treatment $A = 1$ dominates $A = 0$; the reverse is true in the second scenario (upper right). In the third scenario (lower left), the treatments are comparably cost-effective, and the the fourth scenario (lower right), treatment $A = 1$ is moderately more cost-effective as compared to $A = 0$.}
\end{figure}

\subsection{The case for considering both the NMB and NBS}

The NMB and NBS can be seen as complementary measures of cost-effectiveness (the former is a comparison of population means, while the latter is a measure of stochastic ordering). It is possible for the two to provide seemingly contradictory stories at times. In particular, $\nmb(\lambda) > 0 \not\Leftrightarrow \theta(\lambda) > 0.5$. Particularly since it is not clear that one of these measures is uniformly more informative than the other, this provides a case for considering both NMB and NBS in a cost-effectiveness analysis.

Imagine a scenario in which $\theta(\lambda) \approx 0.5$, but $\nmb(\lambda) > 0$. This can happen, for instance, when there is a high level of right-skewness that drives a mean difference. In such a setting, one may put more weight on $\theta(\lambda)$ in regards to decisions on whether to apply the experimental treatment to the entire population. On the other hand, imagine a scenario in which $\theta(\lambda)$ is very high, but that $\nmb(\lambda) \approx 0$. This can happen, for instance, when a large segment of the population exhibits a modest level of treatment benefit with very high precision (i.e., low variability).

These examples, while hypothetical, underscore the tension between scientific questions pertaining to whether a treatment tends to produce more desirable outcomes \textit{on average} and scientific questions pertaining to whether a treatment tends to produce more desirable outcomes \textit{in probability}. Evidence of discordance between $\nmb(\lambda)$ and $\theta(\lambda)$ can spark further exploration in order to learn about possible sources for heterogeneity.

\subsection{The determination curve and its limiting behavior}

The limiting behavior of $\theta(\lambda)$ can be characterized as follows. We have that $\theta(\lambda = 0) = \Pro(Y_0 > Y_1)$ and $\lim_{\lambda \longrightarrow \infty} \theta(\lambda) = \Pro(S_1 > S_0)$, where $(S_a, Y_a)$ denotes a randomly sampled clinical and cost outcome pair under treatment $A = a$. We define the \textit{cost-effectiveness determination} (CED) curve as $\left\lbrace (\lambda, \theta(\lambda)) : \lambda \in \Lambda \right\rbrace$, where again $\Lambda$ denotes a reasonable set of WTP threshold values.

\section{Estimation}

\subsection{Randomized trials with no censoring}

To orient our discussion of estimation, we first consider the idealized setting of a randomized trial in which there is no censoring (this can equivalently be thought of as a randomized trial that considers a non-survival clinical measure). A nonparametric estimate of $\theta(\lambda)$ can be constructed in this case: letting $k = 1, \dots, \sum_{i = 1}^{N} (1 - A_i) \equiv N_0$ index the patients in the control arm and $l = 1, \dots, N_1 \equiv N - N_0$ index the patients in the experimental arm,
\begin{eqnarray*}
	\widehat{\theta}(\lambda) &=& \frac{1}{N_0N_1}\sum_{k = 1}^{N_0}\sum_{l = 1}^{N_1} 1\left(B_{1l}(\lambda) > B_{0k}(\lambda)\right)
\end{eqnarray*}
is trivially unbiased and consistent for $\theta(\lambda)$. It is perhaps better expressed using the equivalent and less computationally taxing scaled variant of the Wilcoxon Rank-Sum statistic:
\begin{eqnarray*}
	\widehat{\theta}(\lambda) &=& \frac{1}{2N_0}\left(\frac{1}{N_1}\sum_{i = 1}^{N} 2A_iR_i(\lambda) - N_1 - 1\right),
\end{eqnarray*}
where $R_i(\lambda)$ denotes the rank of the the pooled INBs. In this setting, the sets $\lbrace B_{0k}(\lambda) \rbrace_{A_k = 0}$ and $\lbrace B_{1l}(\lambda) \rbrace_{A_l = 1}$ each form nonparametric estimates of the marginal distributions for $B_0(\lambda)$ and $B_1(\lambda)$, respectively. The remainder of this section is therefore geared toward approaches for estimating those marginal distributions under more complex (but more realistic) settings in which there is confounding and survival time is the clinical outcome.

\subsection{Semiparametric standardization}

Suppose we are in the setting of observational data. Letting $\bL$ denote (observed) confounders, the NBS can be expressed as:
\begin{eqnarray*}
\theta(\lambda) = \int \theta(\lambda|\bL = \boldsymbol{\ell}) dF_{\boldsymbol{\ell}}(\bL).
\end{eqnarray*}
We may then employ Monte-Carlo standardization for estimation:
\begin{enumerate}
\item Define some range $\Lambda$ of possible willingness-to-pay thresholds.
\item Estimate the distribution of $\bL$ (e.g., via the empirical $\widehat{F}_N(\bL)$).
\item Fit models for $f(S|\bL, A)$ and $f(Y|\bL, A, S)$.
\item For each $a = 0, 1$:
\vspace{-2mm}
\begin{itemize}
\item Generate $K$ random draws $\widetilde{\bL}_k$ from $\widehat{F}_N(\bL)$.
\item Generate $K$ random draws: $\widetilde{S}_k$ from $\widehat{f}(S|\widetilde{\bL}, A = a)$ and $\widetilde{Y}_k$ from $\widehat{f}(Y|\widetilde{\bL}, A = a, \widetilde{S}_k)$, as estimated in Step 3.
\end{itemize}
\vspace{-2mm}
\item Compute the ranks $\widetilde{R}_k(\lambda)$, using the pooled data, of the $K$ INBs from the realizations drawn from $a = 1$.
\item For each $\lambda \in \Lambda$, an estimate the net benefit separation is given by:
\begin{eqnarray*}
\widehat{\theta}(\lambda) &=& \frac{1}{2K}\left(\frac{1}{K}\sum_{k = 1}^{K}2\widetilde{R}_i(\lambda) - K - 1\right).
\end{eqnarray*}
\end{enumerate}
This method specifies parametric models for the conditional distributions of $S$ and $Y$, but not for the distribution of the confounders. Importantly, the parametric models can be quite flexible, easily accommodating regression splines and zero-inflation of cost using a two-part model \cite{Leung1996}.

\subsection{Addressing censoring and truncated cost}
Suppose now that $S$ is survival time. We further assume that $Y$ is considered over time range $[0, \tau]$. Let $C$ denote the censoring time; in practice, $X = \min(S, C)$ is observed. Let $\delta = 1(C < S)$ denote the indicator of censored survival time, and $\delta^* = 1(C < \min(S, \tau))$ the indicator of censored costs. Inverse probability-of-censoring weights (IPCW) must be included to estimate $f(Y|\bL, A)$ owing to informative censoring induced by heterogeneity in cost-accrual rates \cite{spieker2018analyzing, lin2000linear, li2016propensity}. Weights can be obtained via the Kaplan-Meier method. Let $G(t) = \Pro(C \geq t)$, and $\widehat{G}(t)$ the Kaplan-Meier (KM) estimator based on $(X_i, \delta_i^*)$. If covariates are discrete, stratification is possible; for other cases (e.g., continuous covariates), we may instead use employ the stratified Cox proportional hazards model \cite{Cox1972}:
\begin{eqnarray*}
h(t|\bV, \bW) = \exp(\boldsymbol{\phi}^T\bW(t))h_{\bV}(t),
\end{eqnarray*}
where $\bV$ denotes stratification variables and $\bW$ other covariates. Here, $h(\cdot)$ denotes the hazard function for censoring, and we assume that $(S,Y) \ci C|\bV, \bW$; covariates in $\bV$ and $\bW$ may include those in $\bL$. Letting $S^* = \min(S, \tau)$, the probability of being observed can be estimated by:
\begin{eqnarray*}
\widehat{G}_i(S_i^*) &=& \exp\left(-\sum_{j = 1}^{N} \frac{\delta_j 1(\bV_i = \bV_j, S_i^* > X_j)\exp(\widehat{\boldsymbol{\phi}}^T\bW_i(X_j))}{\sum_{k = 1}^{N} 1(\bV_i = \bV_k, X_k \geq X_j)\exp(\widehat{\boldsymbol{\phi}}^T\bW_k(X_j))} \right).
\end{eqnarray*}
Note that $\widehat{\boldsymbol{\phi}}$ denotes the partial likelihood estimator \cite{Cox1975}.

\subsection{Interval estimation and inference}

Confidence intervals can be formed using the nonparametric bootstrap \cite{Davison}. In the case of Section 4.1., $\sqrt{N_1}(\widehat{\theta}(\lambda) - 1/2) \longrightarrow_d \mathcal{N}(0, (r + 1)/(12r))$ under the null hypothesis $H_0 : \theta(\lambda) = 1/2$ and equal variances, where $r$ denotes the randomization ratio $\lim_{N \rightarrow \infty} N_0/N_1$. This may be used for asymptotic inference. Otherwise, hypothesis testing can also be conducted using the nonparametric bootstrap.

\section{Simulation Study}

We conduct a simulation study to evaluate the performance of $\widehat{\theta}(\lambda)$ under two sample sizes ($N = 500$ and $N = $ 5,000) and two censoring rates (10\% and 25\%). Covariates are simulated as $L_1 \sim \mathcal{N}(\mu = 0, \sigma = 1)$ and $L_2 \sim \text{Bernoulli}(0.5)$, and treatment is given by $A \sim \text{Bernoulli}(p = \text{expit}(L_1))$. Survival is generated as $S \sim \text{Weibull}(\text{shape} = 2, \text{ scale} = \exp(\alpha_0 + 0.2 L_1 + 0.2 L_2 + \alpha_A A)$), and censoring is generated as $C \sim \text{Weibull}(\text{shape} = 2, \text{ scale} = \exp(\beta_0 + 0.1 L_2))$. We generate the cost such that $\log(Y) \sim \mathcal{N}(\mu = 4.2 + 0.002 S + \gamma_A A, \sigma = 0.4)$. The parameter $\beta_0$ controls the censoring rate and is set to $\beta_0 = 5.65$ for 10\% censoring (low) and $\beta_0 = 5.1$ for 25\% censoring (high). We consider two simulation scenarios. The first is a null scenario in which the treatment has no impact on cost or survival ($\alpha_0 = 4.5$ and $\alpha_A = \gamma_A = 0$); the second is such that treatment is associated with an increase in both cost and survival ($\alpha_0 = 4.05$, $\alpha_A = 0.7$, and $\gamma_A = 0.1$). We consider $\lambda = 2$ and $\lambda = 12$ as examples.

For each case, we conduct one-thousand replicates with $K = 300$ bootstrap replicates and $M = $ 5,000 Monte-Carlo iterations. The distribution of $(L_1, L_2)$ was estimated nonparametrically using the empirical distribution for the standardization procedure. The Kaplan-Meier estimator is used to determine the IPCW weights, stratified by $L_2$. We report the mean estimate $\widehat{\theta}(\lambda)$ across simulations, as well as the average estimated standard error and empirical standard error across Monte-Carlo replicates.

Tables 1 and 2 present results for each scenario. The proposed procedure is able to estimate $\theta(\lambda)$ with low bias, and the nonparametric bootstrap standard error adequately represents the true repeat-sample variability. For lower samples sizes, the bootstrap standard error slightly underestimates the true standard error on average for lower values of $\lambda$, and slightly overestimate the standard error for higher values of $\lambda$. This bias dissipates with a larger sample size.

\begin{table}[h!]
\caption{Performance of $\widehat{\theta}(\lambda)$ under Scenario 1 (no treatment effect). Included are the mean estimate, empirical standard error (ESE), and mean estimated standard error.}
$$\begin{tabular}{lllcccc}
Censoring & $N$ & $\lambda$ & $\theta(\lambda)$ & Est. & ESE &  $\widehat{\text{SE}}$ \\ \hline
10\% & 500 & 2 & 0.500     & 0.502 & 0.0265 & 0.0230  \\
10\% & 500 & 12 & 0.500    & 0.501 & 0.0260 & 0.0245 \\
10\% & 5,000 & 2 & 0.500   & 0.500 & 0.0120 & 0.0114 \\
10\% & 5,000 & 12 & 0.500  & 0.500 & 0.0119 & 0.0117 \\
25\% & 500 & 2 & 0.500     & 0.500 & 0.0261 & 0.0240 \\
25\% & 500 & 12 & 0.500    & 0.500 & 0.0252 & 0.0257 \\
25\% & 5,000 & 2 & 0.500   & 0.500 & 0.0122 & 0.0117 \\
25\% & 5,000 & 12 & 0.500  & 0.500 & 0.0117 & 0.0120 \\ \hline
\end{tabular}$$
\end{table}

\vspace{6mm}

\begin{table}[h!]
\caption{Performance of $\widehat{\theta}(\lambda)$ under Scenario 2 (treatment increases cost and survival time). Included are the mean estimate, empirical standard error (ESE), and mean estimated standard error.}
$$\begin{tabular}{lllcccc}
Censoring & $N$ & $\lambda$ & $\theta(\lambda)$ & Est. & ESE &  $\widehat{\text{SE}}$ \\ \hline
10\% & 500 & 2 & 0.743     & 0.731 & 0.0242 & 0.0198  \\
10\% & 500 & 12 & 0.780    & 0.771 & 0.0198 & 0.0207 \\
10\% & 5,000 & 2 & 0.743   & 0.730 & 0.0097 & 0.0096 \\
10\% & 5,000 & 12 & 0.780  & 0.771 & 0.0093 & 0.0096 \\
25\% & 500 & 2 & 0.743     & 0.731 & 0.0242 & 0.0216 \\
25\% & 500 & 12 & 0.780    & 0.771 & 0.0203 & 0.0213 \\
25\% & 5,000 & 2 & 0.743   & 0.730 & 0.0110 & 0.0105 \\
25\% & 5,000 & 12 & 0.780  & 0.771 & 0.0099 & 0.0099 \\ \hline
\end{tabular}$$
\end{table}

\section{Application}

We conduct a cost-effectiveness analysis using a simulation-based data set motivated from a prior study of survival and total medical costs in post-hysterectomy stage-I and stage-II endometrial cancer patients \cite{spieker2017causal}. We fit a sequence of flexible models on the variables from the original database and simulated a data set of equal size from the estimated models. Our reasons for taking this approach are twofold: (1) doing so allows us to make the data set fully available; we would not be able to do so otherwise due to the proprietary nature of the data, and (2) the proportion of individuals having an event by the study's end was low due to insufficient follow-up time. Hence, outcomes were simulated over a ten-year window using an aggregate of the original data and estimates from the literature \cite{Shaeffer2005, Susumu2008}. We provide the simulated data set and code in the supplementary materials. Results are intended solely for illustration purposes.

Post-surgery, endometrial cancer patients may receive adjuvant treatment with radiation therapy (RT), chemotherapy (CT), or neither \cite{Braun2016}. Though there are apparent survival advantages associated with adjuvant RT and CT, little has been done to elucidate the cost-effectiveness of either. Our illustration focuses on determining cost-effectiveness associated with each of adjuvant RT alone and adjuvant CT alone received two to four months post-surgery relative to control.

\subsection{Description of data and analysis}
Covariates include age of diagnosis, Charlson co-morbidity index \cite{Charlson1994}, number of prior hospitalizations, cancer stage, and prior receipt of RT or CT. These data include $N = $ 13,526 patients. The mean age of diagnosis was 73.7 years (SD = 6.59 years), and 93.6\% were diagnosed with stage I cancer. At baseline, 5.99\% of patients had received RT and 0.55\% had received CT. Through the two-to-four month window, 25.4\% received RT and 2.25\% received CT (0.79\% received both RT and CT). A total of 7,720 deaths occurred (57.0\%) and 8,762 patients (64.7\%) had complete cost data.

We employ the procedure outlined in Section 4.2 and incorporate weights from the Cox model as outlined in Section 4.3. We utilize the empirical joint distribution of the baseline covariates, and incorporate each covariate as a predictor of censoring time. We fit a parametric Weibull model to the survival distribution conditional on covariates and treatment (each of RT and CT). We estimate the distribution of costs using a two-part zero-inflation model (a logistic model for the probability of zero-costs, followed by a log-normal model for nonzero costs). Quantiles from $K = $ 1,000 bootstrap replicates were used to formulate 95\% confidence intervals, and $M = $ 10,000 replicates were used for the Monte-Carlo draws associated with our standardization procedure.

Our target parameters include $\theta_{R}(\lambda)$ and $\theta_{C}(\lambda)$, which are defined by comparing: (1) RT without CT to control, and (2) CT without RT to control, respectively (where control is defined as no RT or CT). We additionally target $\nmb_R(\lambda)$ and $\nmb_C(\lambda)$, which correspond to the net monetary benefit based on comparisons (1) and (2). We primarily consider a range of 50 to 120 for $\lambda$ (in units of thousands of USD over one year), though we extend the range in order to illustrate limiting behavior. In our analysis, the NMB is determined using the Monte-Carlo draws from the standardization procedure.

\subsection{Results}

Table 3 presents the results of our analysis for the boundary values of $\lambda$ under primary consideration. As an example of how to interpret the estimated NBS, $\widehat{\theta}_C(\lambda)$, consider a WTP threshold of \$120,000, and imagine two hypothetical situations: one in which the population receives CT (but no RT), and another in which the population does not receive adjuvant RT or CT. We estimate that the INB from the individual sampled from the RT population would be higher 67\% of the time upon repeated random sampling from these two hypothetical scenarios, with a 95\% CI of (63\%, 72\%). Figure 3 presents the CED curves for each comparison. Note that $\widehat{\theta}_{R}(0)$ and $\widehat{\theta}_{C}(0)$ are both smaller than 0.5, corresponding to higher costs in the treatment groups. At the lowest value of the WTP used for primary consideration (\$50,000), these results suggest cost-effectiveness for both RT and CT as compared to control.

\begin{table}[h!]
\caption{Point estimates and 95\% confidence intervals for parameters of interest (columns) for the lower and upper values of interest for the WTP (rows). Confidence intervals are based on the 2.5$^\text{th}$ and 97.5$^\text{th}$ quantiles of the nonparametric bootstrap. Note that $\lambda$ and NMB are in units of thousands of dollars.}
$$\begin{tabular}{lcccc}
$\lambda$ & ${\theta}_{R}$ & ${\theta}_{C}$ & $\text{NMB}_{R}$ & $\text{NMB}_{C}$ \\ \hline
50  & 0.59 (0.58, 0.61) & 0.63 (0.60, 0.69) & 46  (38, 69) & 70  (26, 126) \\
120 & 0.60 (0.59, 0.63) & 0.67 (0.63, 0.72) & 141 (127, 182) & 261 (185, 381) \\ \hline
\end{tabular}$$
\end{table}

\clearpage

Table 3 additionally presents the estimated NMB at the same values of $\lambda$ along with 95\% confidence intervals. Figure 4 depicts the estimated NMB, with confidence intervals included for the primary range of $\lambda$. Similarly to the NBS, the confidence intervals do not include zero, suggesting evidence of cost-effectiveness as measured by mean differences. Figure 5 depicts corresponding CEA curves, which suggest strong evidence that both RT and CT are cost-effective relative to control on average for $\lambda > 50$. Moreover, the strength of evidence is not sensitive to changes in $\lambda$ over the range of interest (note this would not be so for lower values of $\lambda$). These insights could also be gleaned from examination of Figure 4.

In this example, we find that the results from the NBS are consistent with those of the NMB, suggesting that both RT and CT are cost-effective relative to control.

\begin{figure}[h!]
    \centering
    \includegraphics[width = 4.4in]{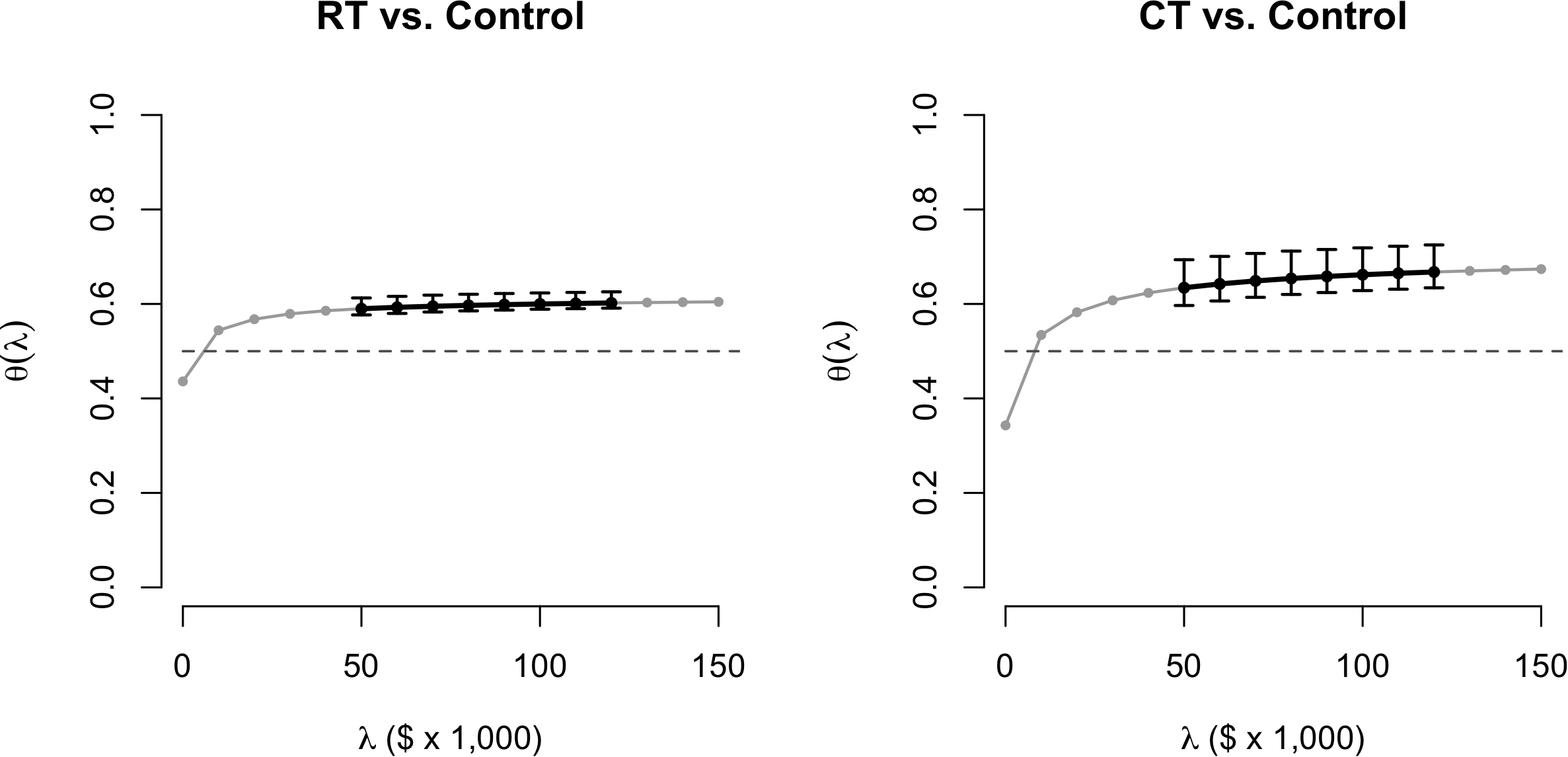}
    \caption{Cost-effectiveness determination curve for RT (left) and CT (right) vs. control over the willingness-to-pay range of interest (bold). Corresponding point-wise confidence intervals are included. For the purposes of illustrating the scope of curve behavior, we extend the range of $\lambda$ beyond the range of interest (shown in lighter gray).}
\end{figure}

\begin{figure}[h!]
    \centering
    \includegraphics[width = 4.4in]{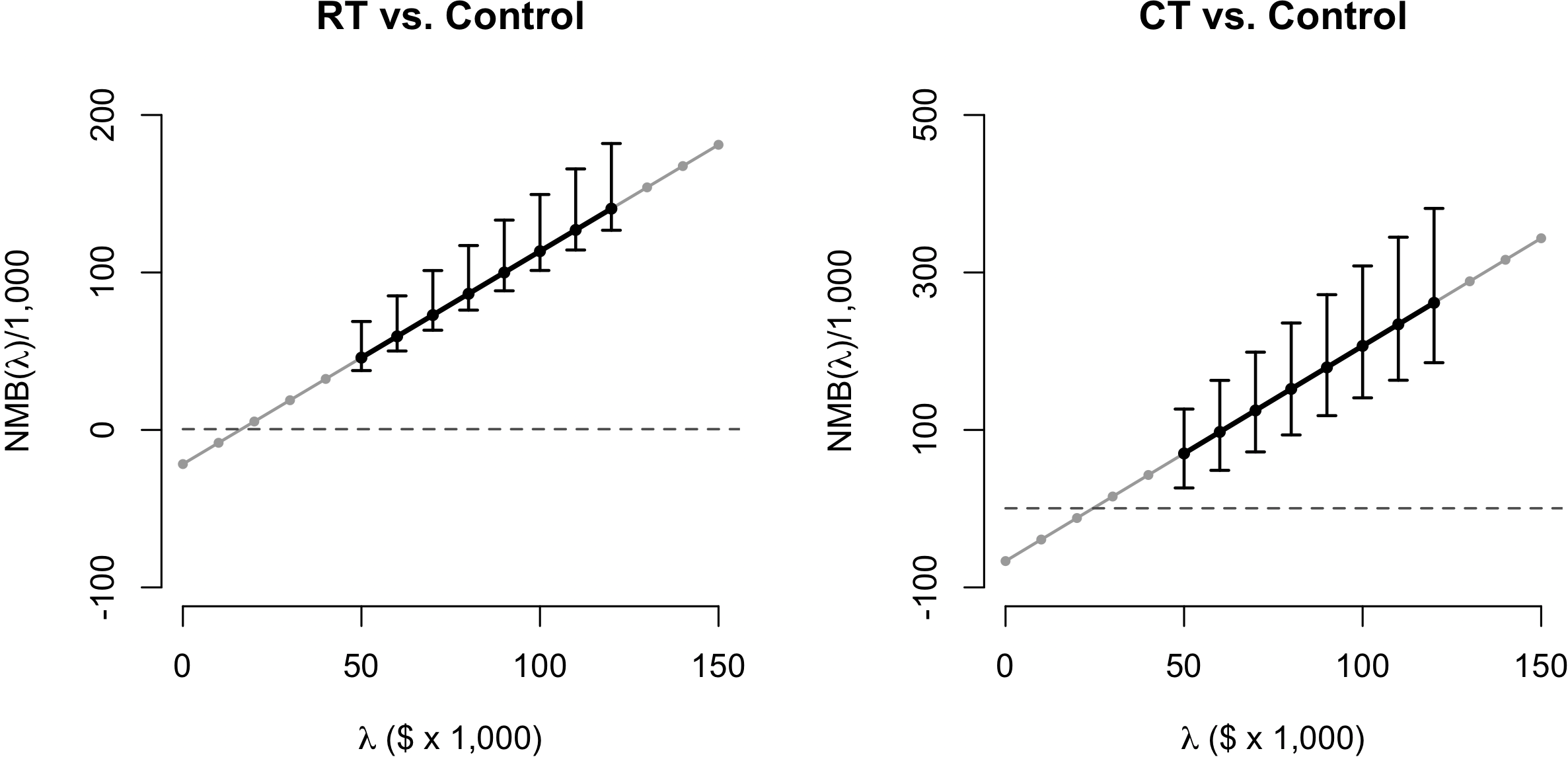}
    \caption{Plot of the NMB for RT (left) and CT (right) vs. control over the willingness-to-pay range of interest (bold). For the purposes of illustrating the scope of curve behavior, we extend the range of $\lambda$ beyond the range of interest (shown in lighter gray).}
\end{figure}

\begin{figure}[h!]
    \centering
    \includegraphics[width = 4.4in]{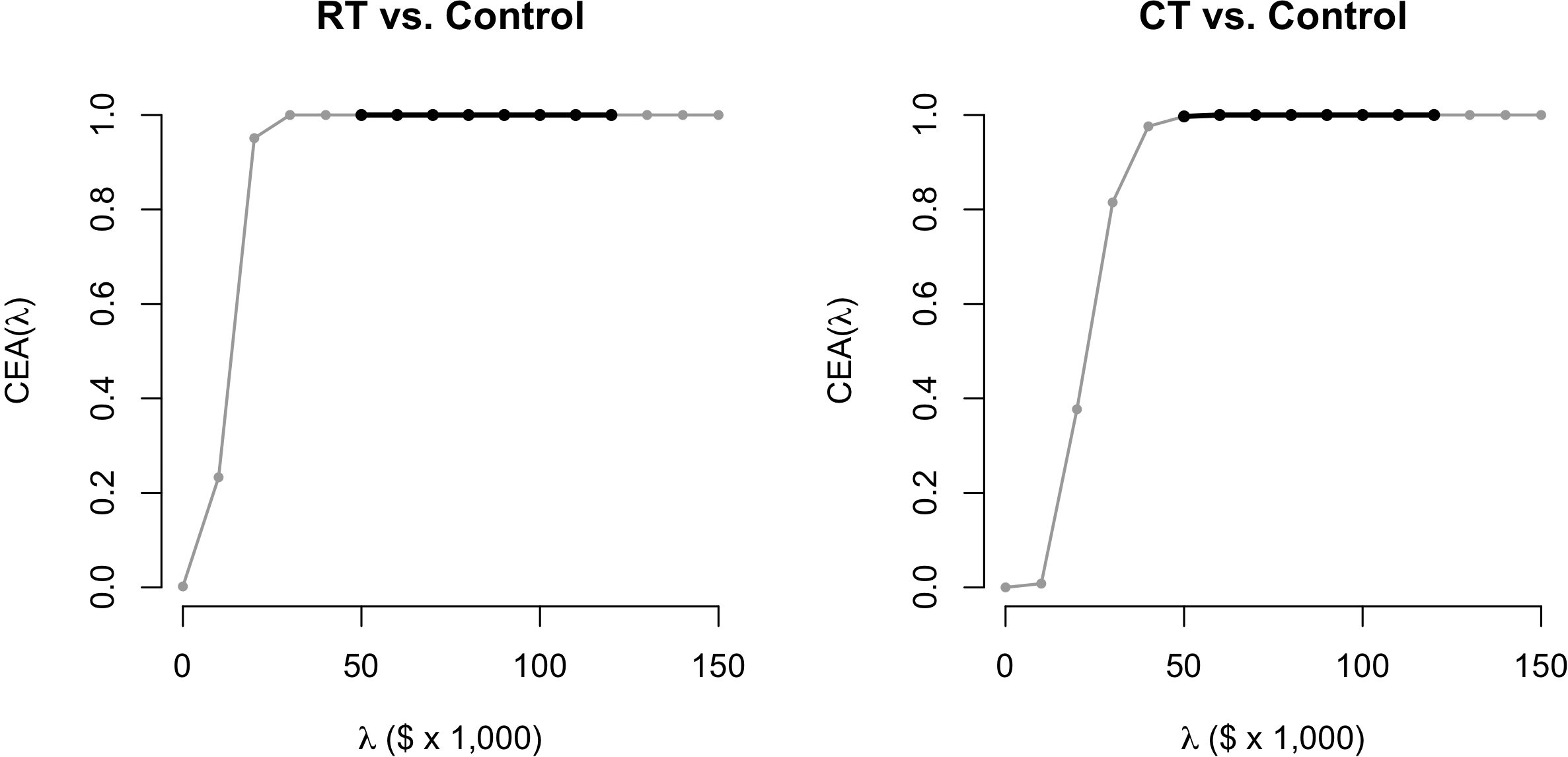}
    \caption{Cost-effectiveness acceptability curve for RT (left) and CT (right) vs. control over the willingness-to-pay range of interest (bold). For the purposes of illustrating the scope of curve behavior, we extend the range of $\lambda$ beyond the range of interest (shown in lighter gray).}
\end{figure}

\section{Discussion}

In this paper, we have first summarized the net monetary benefit and its corresponding acceptability curve, both of which are standard in cost-effectiveness analysis. We stated the proper interpretation of results from these approaches and underscored some of the limitations associated with current methodology.

Our proposed novel metric, the NBS, is a probabilistic population parameter that quantifies the magnitude of cost-effectiveness, having an interpretation similar in spirit to that of the area under a receiver operating characteristic curve in diagnostic testing. Our approach to estimation can accommodate challenges typically encountered in observational studies such as confounding and censoring via standardization and inverse probability weighting, respectively.

The limitations of the acceptability curve that we have delineated do not generally apply to the net monetary benefit as a whole. The net monetary benefit is a summary of average cost-effectiveness and is useful in its own right. In order to graphically understand how the magnitude of the net monetary benefit and evidence of a non-zero average effect on cost-effectiveness change across the willingness-to-pay, we demonstrate that one may plot the estimated net monetary benefits and respective confidence intervals across values of $\lambda$ with no need to consider the acceptability curve.

In Section 3.1, we describe scenarios in which the NMB and NBS could provide seemingly contradicting information. This potential dichotomy cannot be resolved without fully understanding the context of the treatment and disease setting, including each treatment's safety profile and the clinical outcomes under consideration. It also certainly cannot be resolved unless both measures are computed and understood. A discordance between the two metrics would likely prompt subgroup analyses based on important variables to try to reconcile the systematic sources of heterogeneity; we are currently developing approaches for subgroup analysis in cost-effectiveness.

One potential advantage of the net monetary benefit is the well-ordering/transitivity property associated with simultaneous comparison of multiple competing treatment regimes. That is, if treatment $A$ is better than treatment $B$ and treatment $B$ is better than treatment $C$, then treatment $A$ will be better than treatment $C$ when "better" means a higher NMB. It is known that transitivity does not always hold with measures of stochastic ordering like the NBS. However, the NBS satisfies the previously unaddressed methodological gap pertaining to the lack of a sensible probabilistic measure of cost-effectiveness magnitude.

Issues surrounding unobserved confounding apply to both the NMB and the NBS approach, and always merits close scrutiny in observational data. The risk of bias due to unobserved confounding is always real in observational data, although care can be taken in many cases at the data collection stage in order to to mitigate the impact. Approaches to assess sensitivity of results to the assumption of no unmeasured confounding would be of interest in this setting \cite{Handorf2013}.

Our application is designed to illustrate a properly conducted cost-effectiveness analysis---one in which insights regarding both the net monetary benefit and net benefit separation are aggregated to paint a complete picture regarding treatment's cost-effectiveness. Importantly, our method for determining the NMB in this example is distinct from that employed in prior literature. Li et al. \cite{jiaqili2018} propose IPCW and propensity-score weights in order to estimate the NMB; this approach is based on restricted mean survival time, in which the restriction is to the upper bound on censoring time. Since our parametric approach for estimating the survival distribution does not demand restriction, considering the semi-parametric approach of Li et al. would induce an incongruity in a comparison of the two cost-effectiveness methods. 

\clearpage

{\small \singlespacing

}

\end{document}